\documentclass[preprint,12pt]{elsarticle}

\usepackage{epsfig}
\usepackage{multirow}
\usepackage{amssymb}

\journal{Annals of Physics}

\begin{document}

\begin{frontmatter}



\title{Schemes to avoid entanglement sudden death of decohering two qubit system}


\author{Namitha C V and S V M Satyanarayana}

\address{Department of Physics, Pondicherry University, Puducherry 605 014, India}
\ead{svmsatya@gmail.com}

\begin{abstract}
We investigate the entanglement dynamics of two interacting qubits in a common vacuum environment. The inevitable environment interaction leads to entanglement sudden death (ESD) in a two qubit entangled state system. The entanglement dynamics can be modified by the use of local unitary operations (quantum gates), applied on the system during its evolution. We show that these operations not only delays or avoids the ESD but also advances the entanglement revival with high concurrence value depending on the time of operation. We have analytically found out different time windows for switching with different quantum gates so that the ESD can be completely avoided in the subsequent evolution of the system. Our result offers practical applications in the field of quantum information processing where the entanglement is a necessary resource.
\end{abstract}


\end{frontmatter}


\section{Introduction}
\label{}
Quantum entanglement is an important quantum correlation that has been playing a central role in quantum information and communication theory. Entangled states are important in many applications such as quantum teleportation, quantum computation and quantum key distribution \cite{Nielsen2000,Bennett1993,Horodecki2009,Brunner2014}. A bipartite quantum system, in addition to mutual interaction, also interacts with surrounding environment in open quantum systems. Initially entangled coherent bipartite system upon evolution in noisy environment, undergoes decoherence and degradation of entanglement \cite{Leandro2015}. The decoherence can also lead to an abrupt loss of entanglement which is termed as entanglement sudden death (ESD) \cite{Yu2004,Yu2006,Yu2009}. ESD has been confirmed experimentally \cite{Almeida2007,Laurat2007}. Work has been done on entanglement dynamics of two qubit system where creation of entanglement \cite{Ficek2002,Ficek2003,RTanas2004,ZFicek2010} sudden birth \cite{Ficek2008} and revival \cite{Ficek2006} of entanglement are observed during the decoherence process of a two qubit system. While high amount of entanglement of a bipartite system that can sustain for long time is a requirement for many applications in quantum information science, such as quantum teleportation, quantum key distribution and quantum computation, occurence of ESD induced by decoherence poses a serious threat to the success of these applications. Thus it is essential to develop schemes that can avoid or delay ESD. Different methods available in the literature are based on the use of Decoherence free subspace \cite{Lidar2006,Kwiat2000}, Quantum Zeno effect \cite{Facchi2004,Maniscalco2008}, Quantum error correction \cite{Shor1995,Steane1996}, use of local unitary operations \cite{Anna2006,Yu2007}, direct quantum feedback \cite{Carvalho2008}, quantum interference \cite{Sumanta2010}, dynamical decoupling ~\cite{Lorenza1999, Pan2011}, delayed choice decoherence suppression~\cite{Jong2014} and weak measurement and quantum measurement reversal \cite{Kim2012}. Among these, use of local unitary operation is a simple method and received broad attention. Depending on the time of operation, local unitary operators can delay ESD or avoid ESD ~\cite{Yu2007,Rau2008,Mahmood2012}. An analytical expression for controlling ESD in X states by use of local unitary operations in a statistically independent thermal reservoir \cite{Ali2009} has been presented and proved that if atleast one reservoir is at finite temperature all X states exhibits ESD which cannot be avoided. Also, switching the environment from thermal to vacuum can delay the ESD \cite{Mahmood2011}. In the present work we consider two identical two level atoms with mutual dipole interaction that interact with a common vacuum bath. Here, decoherence results from spontaneous emission. In the evolution of a state under Markov approximation, we develop schemes based on local unitary operation to completely avoid ESD in an important class of initial entangled states. We also propose different time windows for switching for a class of X states with one photon and two photon coherence properties.
\subsection{Two atom model}
The theoretical model composed of two identical atoms, which are spatially separated, interact with each other through dipole forces and coupled to a common vacuum field. The atoms are modeled as two level systems with ground states $\vert g_{i}\rangle$ and excited states $\vert e_{i}\rangle$ $(i=1,2)$ which form the computational basis for the atomic system. The interaction of two level atomic system with environment results in spontaneous emission and the dynamics can be studied using the Lehemberg-Agarwal master equation under Markovian approximation~\cite{Lehemberg1970, Ficek1987}. The time evolution of the two qubit system interacting with a vacuum bath is given by the following master equation
\begin{equation}\label{ME}
\frac{\rm d \rho}{\rm d t}=- {\rm i} \sum_{i=1}^{2}\omega_{i}[S_{i}^{z},\rho]- {\rm i} \Omega_{ij}\sum_{i\neq j}^{2}[S^{+}_{i}S^{-}_{j},\rho] -\frac{1}{2}\sum_{i,j=1}^{2}\Gamma_{ij}(\rho S_{i}^{+} S_{j}^{-}+ S^{+}_{i} S^{-}_{j}\rho - 2 S^{-}_{j}\rho S^{+}_{i}),
\end{equation}
where $S^{+}_{i}$, $S^{-}_{i}$ and $S^{z}_{i}$ are respectively the dipole raising, lowering and energy operators of the $i^{th}$ atomic qubit. The transition frequencies of atoms are same since the atoms are identical, which is taken as $\omega_{i=1,2}=\omega_{0}$. And $\Gamma_{ij}$ for ($i=j$) is equal to $\Gamma$ are the spontaneous emission rates of the qubits. $\Gamma_{ij}$ and $\Omega_{ij}(i\neq j)$ represent the emission rates and the interatomic coupling parameters respectively and are given by
\begin{equation}\label{Gamma}
\Gamma_{ij}=\Gamma_{ji}=\frac{3}{2}\Gamma\{[1-(\hat{\mu}.\hat{r}_{ij})^{2}]\frac{\sin(kr_{ij})}{kr_{ij}}
+[1-3(\hat{\mu}.\hat{r}_{ij})^{2}][\frac{\cos(kr_{ij})}{(k r_{ij})^2}-\frac{\sin(kr_{ij})}{(kr_{ij})^3}]\},
\end{equation}

\begin{equation}\label{Omega}
\Omega_{ij}=\frac{3}{4}\Gamma\{-[1-(\hat{\mu}.\hat{r}_{ij})^{2}]\frac{\cos(kr_{ij})}{kr_{ij}}
+[1-3(\hat{\mu}.\hat{r}_{ij})^{2}][\frac{\sin(kr_{ij})}{(k r_{ij})^2}+\frac{\cos(kr_{ij})}{(kr_{ij})^3}]\},
\end{equation}
where $\hat{\mu}$ is the unit vector along the direction of dipole moment of the atoms, $\hat{r}_{ij}$ is the unit vector along the direction of $\vec{r}_{ij}$ and $k=\frac{\omega_0} {c}$.

The energy eigenstates of an isolated non interacting atomic system is the product basis of the composite Hilbert space. However, coupling to the environment the system behaves collectively as a single four-level system. The collective atomic basis is called as Dicke state basis~\cite{Dicke1954}, defined as

\begin{eqnarray}
 \vert g\rangle &=& \vert g_{1} g_{2}\rangle, \\
  \vert s\rangle &=& \frac{1}{\sqrt{2}}[\vert e_{1}\rangle \vert g_{2}\rangle+\vert g_{1}\rangle  \vert e_{2}\rangle ], \\
  \vert a\rangle &=& \frac{1}{\sqrt{2}}[\vert e_{1}\rangle \vert g_{2}\rangle-\vert g_{1}\rangle  \vert e_{2}\rangle ], \\
  \vert e\rangle &=& \vert e_{1} e_{2}\rangle.
\end{eqnarray}

The collective atomic states consist of upper state $\vert e_{1} e_{2}\rangle$, ground state $\vert g_{1} g_{2}\rangle$ (separable states) and two intermediate maximally entangled states; symmetric $ \vert s\rangle$ and antisymmetric $\vert a\rangle$ state.
\subsection{Measure of Entanglement}

We use concurrence to quantify the entanglement of the two qubit system and employ Wootter's analytical expression for concurrence ~\cite{Wootter1998}, given by

\begin{equation}\label{C}
C={\rm max} \{ 0,\sqrt{\lambda_{1}}-\sqrt{\lambda_{2}}-\sqrt{\lambda_{3}}-\sqrt{\lambda_{4}}\},
\end{equation}
where $\lambda^{'s}$ are the eigenvalues of $\rho\rho\tilde{}$ in descending order. The spin flipped density matrix $\rho\tilde{}$ can be written as

\begin{equation}
\rho\tilde{}=\sigma_{y}^{A}\otimes\sigma_{y}^{B}\rho^{*}\sigma_{y}^{A}\otimes\sigma_{y}^{B},
\end{equation}
where $\rho^{*}$ is the complex conjugate of the density matrix $\rho$. C is zero for separable states and $ 0<C\leq1 $ for entangled states.
In this work we consider the initial states that belong to the class of X states which has the following form as

\begin{equation}
\left(
  \begin{array}{cccc}
    \rho_{11} & 0 & 0 & \rho_{14} \\
    0 & \rho_{22} & \rho_{23} & 0 \\
    0 & \rho_{32} & \rho_{33} & 0 \\
    \rho_{41} & 0 & 0 & \rho_{44} \\
  \end{array}
\right).
\end{equation}

The time evolution preserves the X form of the state. The concurrence of a density matrix in X form in the computational basis ~\cite{Yu2005,Jakóbczyk2004} gets simplified as

\begin{equation}\label{c}
  C(t)={\rm max} \{ 0,C_{1}(t),C_{2}(t)\},
\end{equation}

where

\begin{equation}\label{c1}
  C_{1}(t)=2[\vert\rho_{14}(t)\vert-\sqrt{\rho_{22}(t)\rho_{33}(t)}],
\end{equation}

and

\begin{equation}\label{c2}
  C_{2}(t)=2[\vert\rho_{23}(t)\vert-\sqrt{\rho_{11}(t)\rho_{44}(t)}].
\end{equation}
The given X states are entangled if and only if density matrix satisfies the condition of either $\rho_{22}(t)\rho_{33}(t)< \vert\rho_{14}(t)\vert^{2}$ or $\rho_{11}(t)\rho_{44}(t)< \vert\rho_{23}(t)\vert^{2}$. The expression for concurrence in collective basis is given by

\begin{equation}\label{C1}
  C_{1}(t)=2\vert\rho_{ge}(t)\vert-\{[\rho_{ss}(t)+\rho_{aa}(t)]^{2}-[2 Re \rho_{sa}(t)]^{2}\}^{1/2},
\end{equation}

and

\begin{equation}\label{C2}
  C_{2}(t)=\{[\rho_{ss}(t)-\rho_{aa}(t)]^{2}+[2 Im \rho_{sa}(t)]^{2}\}^{1/2}-2 \sqrt{\rho_{ee}(t)\rho_{gg}(t)}.
\end{equation}
Through the expressions for $C_{1}(t)$ and $C_{2}(t)$, the entanglement is seen to depend on two-photon coherence term,  $\rho_{14}$ and one-photon coherence term, $\rho_{23}$ ~\cite{Ficek2010}.

\section{ESD and Control of ESD using local unitary operators}
ESD \cite{Yu2004,Yu2006,Yu2009,Ficek2006,Shaukat2018} is a serious limiting factor in the real experiments involving entangled state as a necessary resource. Among several methods to control ESD, use of local unitary operation during the evolution is simple and promising to control the decoherence ~\cite{Yu2005,Anna2006,Rau2008}. All-optical experimental set up has been proposed ~\cite{Ashutosh2017} based on local NOT operations that can hasten, delay or avoid ESD in a photonic system undergoing decoherence via amplitude damping channel. The action of local unitary operators on a quantum state at a particular time during the evolution does not alter the entanglement and this action is also termed as switching. In the present work we deal with a method which consists of the use of local unitary operations that improve entanglement longevity of a two qubit system when it undergoes decoherence. We choose different initial entangled states that belong to a class of X states and investigate entanglement dynamics of these states when local unitary operations implemented.
The action of local unitary operations on the system is implemented by quantum gate operations. A general $2\times2$ unitary matrix acting on a single qubit system is defined as
\begin{equation}\label{U}
U(\theta,\phi)=\left({\begin{array}{cc}
 \rm e^{-i \phi}\cos \frac{\theta}{2} & \sin \frac{\theta}{2} \\
 -\sin \frac{\theta}{2} & \rm e^{i \phi} \cos \frac{\theta}{2}  \\
     \end{array}}
\right),
\end{equation}
where $ 0 \leq \theta \leq \pi$ and $0 \leq \phi \leq \frac{\pi}{2}$. By choosing  different values for $\theta$ and $\phi$, we can construct various quantum gates. For example, $U(0,\pi)$ corresponds to Pauli-Z gate operator and $\lbrace U(0,\pi).U(\pi,0) \rbrace $ form Pauli-X gate operator and so on. The commonly used quantum gates operations are Pauli-X gate and Pauli-Z gate. A given initial state $\rho(0)$ evolves under Markov approximation to $\rho(t_{s})$, where $t_{s}$ is the time of switching. Upon switching the state is transformed as

\begin{equation}\label{U1}
\rho'(t_{s})=(\sigma^{(1)}_{i}\otimes\sigma^{(2)}_{j}) \rho(t_{s}) (\sigma^{(1)}_{i}\otimes\sigma^{(2)}_{j})^{\dagger},
\end{equation}
where $\sigma^{(1)}_{i}$ and $\sigma^{(2)}_{j}$ are local unitary operators of first and second systems of a bipartite system. After switching, the state $\rho'(t_{s})$ evolve under same Markov approximation to $\rho'(t)$ for $t>t_{s}$. In our study we use Pauli X-gate, Pauli Z-gate and their different combinations as quantum gate operations.

In what follows, we systematically investigate the implementation of X-I, Z-I and X-Z gates on a select but important class of quantum states. Our aim is to find a gate operation and a time window for switching such that ESD is completely avoided.

\subsection{Bell states}
\begin{figure}[h]
\center
\includegraphics[width=15cm, height=10cm]{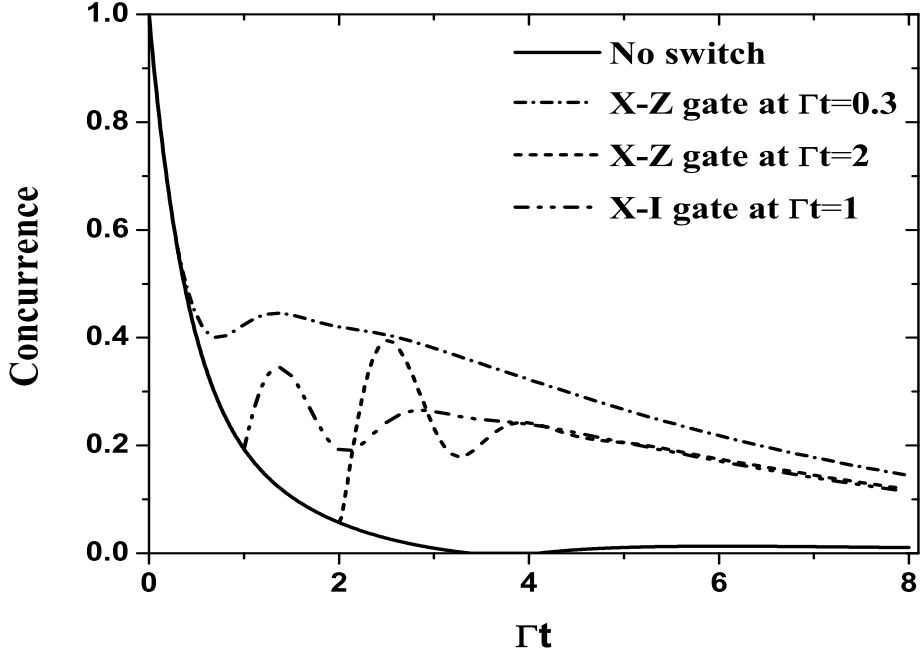}
\caption{Variation of concurrence with time for the evolution of $|\Phi^{+}\rangle$ state on switching with different gate operations, $r_{12}=\lambda/6$ ($\Gamma_{12}=0.79\Gamma$, $\Omega_{12}=1.12\Gamma$)}
\label{fig:1}       
\end{figure}

Bell states are maximally entangled pure states which are linear superpositions of product states and can not be separated in to product states of individual atoms. It is written as
\begin{equation}\label{Psi}
|\Psi ^{\pm}\rangle= \frac{1}{\sqrt{2}}[\vert 0 1 \rangle \pm \vert 1 0 \rangle ],
\end{equation}

\begin{equation}\label{Phi}
|\Phi ^{\pm}\rangle= \frac{1}{\sqrt{2}}[\vert 0 0 \rangle \pm \vert 1 1 \rangle ].
\end{equation}
The nonlocal properties of all Bell states are same. However, they behave very differently during their entanglement evolution. That is, if we choose initial state to be $|\Psi^{-}\rangle$, there is no ESD and the entanglement evolves with the population of antisymmetric state which decays on a much longer time scale. If the chosen initial state is, $|\Psi^{+}\rangle$ then it decays asymptotically with the population of symmetric state. This is because if the initial state is $|\Psi^{-}\rangle$ ($|\Psi^{+}\rangle$), the spontaneous emission results from $|\Psi^{-}\rangle$ ($|\Psi^{+}\rangle$) to the ground state $|g>$ and hence entanglement evolution coincides with that of respective population of antisymmetric (symmetric) state. On the other hand $|\Phi^{\pm}\rangle$ states exhibit ESD followed by revival of entanglement ~\cite{Ficek2006,Ficek2010,ZFicek2010}. Let $\Gamma$t$_{D}=\tau_{D}$ and $\Gamma$t$_{R}=\tau_{R}$ respectively denotes the time at which state undergoes ESD and entanglement revival. Then for $|\Phi^{\pm}\rangle$ states, ESD time, $\tau_D = 3.3883$ and revival time, $\tau_R=4.0972 $ as given in table 1. In this case spontaneous emission connects excited state to symmetric and antisymmetric states along with symmetric and antisymmetric to the ground state.
\begin{table}[h!]
\caption{\label{tabone}ESD time and revival time of different Bell states.}
\begin{center}
\begin{tabular}{c c c} 
\hline\hline
Bell state & ESD time ($\tau_D$) & Revival time ($\tau_R$) \\
\hline
$ |\psi ^ + \rangle$ & No ESD &  ---  \\
$ |\psi ^ - \rangle$ & No ESD & --- \\
$ |\Phi ^ \pm \rangle$ & 3.3883 & 4.0972 \\
\hline\hline
\end{tabular}
\end{center}
\end{table}
From table 1 it is seen that only Bell state (\ref{Phi}) undergoes ESD followed by a revival during the evolution. We investigate the effect of different local unitary operations, namely, X-I, Z-I and X-Z gates on the entanglement dynamics of $|\Phi^{+}\rangle$ state. We observe that the use of X-I gate operation from $0 < \Gamma t < 0.68$, advances the ESD and from $0.68 \leq \Gamma t \leq 2$ avoids ESD and further switching from $2 \leq \Gamma t \leq \tau_D$ again advances the ESD. Next, the action of Z-I gate switching between $0 < \Gamma t \leq \tau_D$ hastens the ESD and thus Z-I gate can not be used to reduce the ESD for $|\Phi^{+}\rangle$ state and the concurrence evolution is dominated by both (\ref{C1}) and (\ref{C2}). Further, we use X-Z gate gate operation and switching from $0 \leq \Gamma t  \leq 0.62$  and  $1.4 < \Gamma t \leq \tau_D$ avoids ESD. However, switching from $0.62 < \Gamma t \leq 1.4$ advances the ESD. It is given in table 2. Thus, we can say that to avoid ESD for initial $|\Phi^{+}\rangle$ state, we have to switch with either X-I gate at $0.68 \leq \Gamma t \leq 2$ or  X-Z gate at $0 \leq \Gamma t \leq 0.62$  and at $1.4 < \Gamma t \leq \tau_D$. It is also important to note that if the switch is initiated after the ESD time, it advances the revival time and make to evolve further with high concurrence. We have plotted the variation of concurrence for the switching with different gates that avoids ESD in $|\Phi^{+}\rangle$ state and is shown in figure~\ref{fig:1}. From our investigation we have seen that the switching action of X-I and X-Z gates on $|\Phi^{+}\rangle$ is interchanged in the case of $|\Phi^{-}\rangle$ and the results are summarized in table 2 and table 3.

\begin{table}[h!]
\begin{center}
\caption{\label{tabone}Switching action in $|\Phi^{\pm}\rangle$ states.}
\begin{tabular}{c c c c c}
\hline \hline
 Initial states & Quantum gates & Avoids ESD & Delays ESD & Hastens ESD \\
\hline
                    & X-I &  $0.68\leq \Gamma t \leq 2$  & --- & \begin{tabular}{c} $0< \Gamma t <0.68;$ \\ $2< \Gamma t \leq \tau_D$ \\
\end{tabular}  \\

$|\Phi^{+}\rangle$ & Z-I & ---  &  ---  & $0 < \Gamma t \leq \tau_D$   \\
                    & X-Z & \begin{tabular}{c} $0 < \Gamma t \leq 0.62;$ \\ $1.4< \Gamma t \leq \tau_D$ \\
\end{tabular}  & --- & $0.62< \Gamma t \leq 1.4$ \\
\hline
                   & X-I & \begin{tabular}{c} $0 < \Gamma t \leq 0.62;$ \\ $1.4< \Gamma t \leq \tau_D$ \\
\end{tabular}  & --- & $0.62< \Gamma t \leq 1.4$ \\

$|\Phi^{-}\rangle$ & Z-I & ---  &  ---  & $0 < \Gamma t \leq \tau_D$   \\
                    & X-Z & $0.68\leq \Gamma t \leq 2$  & --- & \begin{tabular}{c} $0< \Gamma t <0.68;$ \\ $2< \Gamma t \leq \tau_D$ \\
\end{tabular}  \\

\hline\hline
\end{tabular}
\end{center}
\end{table}

\begin{table}[h!]
\caption{\label{tabone}Scheme for avoiding ESD in different Bell states.}
\begin{center}
\begin{tabular}{c c c}
\hline \hline
 Bell state & Time interval & Quantum gate \\
\hline
\multirow{2}{*}{$ |\Phi ^ + \rangle$} & \multicolumn{1}{l}{(0,0.62]; (1.4,$\tau_D$]} & \multicolumn{1}{l}{X-Z} \\
                                      & \multicolumn{1}{l}{[0.68,2]} & \multicolumn{1}{l}{X-I} \\

\hline
\multirow{2}{*}{$ |\Phi ^ - \rangle$} & \multicolumn{1}{l}{(0,0.62]; (1.4,$\tau_D$]} & \multicolumn{1}{l}{X-I} \\
                                      & \multicolumn{1}{l}{[0.68,2]} & \multicolumn{1}{l}{X-Z} \\
\hline \hline
\end{tabular}
\end{center}
\end{table}

\subsection{Werner states}
Pure states are considered to be ideal states but they evolve into mixed states when they undergo decoherence. Mixed states are very important quantum states since they correspond to real field experiments. Let us consider the important class of mixed quantum states, Werner state is a convex sum of maximally entangled pure state and maximally mixed separable state~\cite{Werner1989}. Werner state is defined as

\begin{equation}\label{W}
\rho_{W}^{M}=(\frac{1-p}{4}){I_{4}}+ p |M\rangle \langle M |,
\end{equation}
with $0\leq p \leq1$ and $| M \rangle$ can be any one of the Bell states.

Both the Bell states and Werner states belong to the class of X states. The non-local properties of all four Werner state are identical, that is the states are entangled for $\frac{1}{3} < p \le 1$ and states violate Bell CHSH inequality for $p > \frac{1}{\sqrt{2}}$. However, if we consider the Werner state as an initial state, the entanglement evolution is not identical~\cite{Anna2006}. If the initial state is $\rho_{W}^{\Psi^{-}}$, there is no ESD. For a fixed mixedness, say, the linear entropy $S_{L}(0)=0.7$, Werner states $\rho_{W}^{\Psi^{+}}$ and $\rho_{W}^{\Phi^{\pm}}$ exhibit ESD followed by revival of entanglement. The ESD time and revival time of Werner state for a fixed mixedness $S_{L}(0)=0.7$ and $p=0.5477$ is given in table 4. It is observed that ESD occurs faster if the initial state is $\rho_{W}^{\Psi^{+}}$  compared to $\rho_{W}^{\Phi^{\pm}}$ state. For all Werner states, the ESD time and revival time is found to increase with increase in initial entanglement. Also, we have studied the dependence of $\tau_{D}$ and $\tau_{R}$ on initial mixedness, which is found to decreases with increase in linear entropy. That is, higher the initial purity of the Werner state higher the ESD time and revival time.

\begin{table}[h!]
\caption{\label{tabone}ESD time and revival time of a representative Werner states with $S_{L}(0)=0.7$ and $p=0.5477$.}
\begin{center}
\begin{tabular}{c c c}
\hline\hline
 Werner state & ESD time ($\tau_D$) & Revival time ($\tau_R$) \\
\hline
$\rho_{W}^{\Psi^{-}}$ & No ESD &  ---  \\
$\rho_{W}^{\Psi^{+}}$ & 0.2871 &  2.3428 \\
$\rho_{W}^{\Phi^{\pm}}$ & 0.613 & 2.7227 \\
\hline\hline
\end{tabular}
\end{center}
\end{table}

\begin{figure}[h]
\center
\includegraphics[width=15cm, height=10cm]{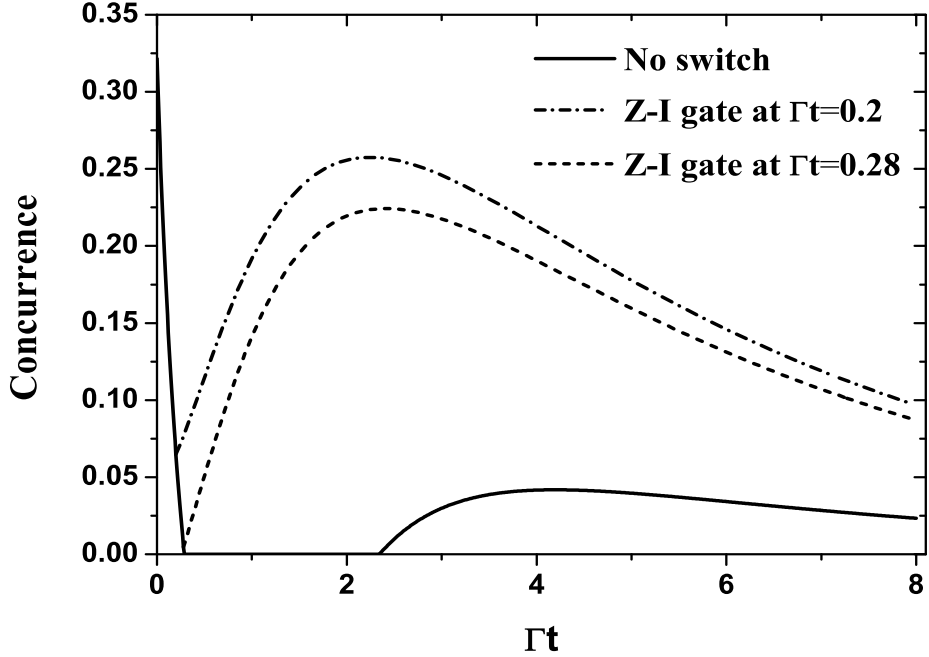}
\caption{Variation of concurrence with time for the evolution of  $\rho_{W}^{\Psi^{+}}$ state before (solid black) and after switching with Z-I gate at $\Gamma t =0.2$ (dash dotted red) and $\Gamma t =0.28$ (dashed blue) for $S_{L}(0)=0.7$, $r_{12}=\lambda/6$ ($\Gamma_{12}=0.79\Gamma$, $\Omega_{12}=1.12\Gamma$)}
\label{fig:2}       
\end{figure}

\begin{figure}[!h]
\center
\includegraphics[width=15cm, height=10cm]{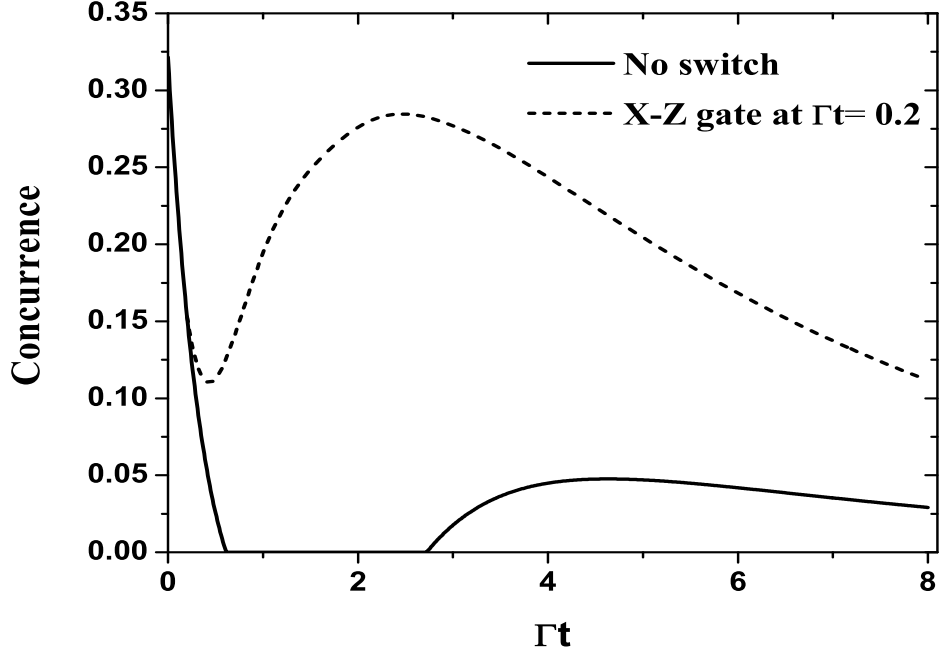}
\caption{Concurrence evolution of $\rho_{W}^{\Phi^{+}}$ state before (solid black) and after switching with X-Z gate at $\Gamma t =0.2$ (dashed red) for $S_{L}(0)=0.7$, $r_{12}=\lambda/6$ ($\Gamma_{12}=0.79\Gamma$, $\Omega_{12}=1.12\Gamma$)}
\label{fig:3}       
\end{figure}

We investigate the effect of different local unitary operations, namely, X-I, Z-I and X-Z gates on the entanglement dynamics of initial Werner states. For example, we have seen from table 4 that Werner states $\rho_{W}^{\Psi^{+}}$ and  $\rho_{W}^{\Phi^{\pm}}$ exhibits ESD and revival of entanglement. Results of switching on these three states are given in table 5.
\begin{table}[h!]
\caption{\label{tabone}Switching action in Werner states with $S_{L}(0)=0.7$ and $p=0.5477$.}
\begin{center}
\begin{tabular}{c c c c c}
\hline\hline
 Initial states & Quantum gates & Avoids ESD & Delays ESD & Hastens ESD \\
\hline
                    & X-I &  ---  & $0< \Gamma t < 0.25$ & $0.25 \leq \Gamma t \leq \tau_D$  \\
$\rho_{W}^{\Psi^{+}}$ & Z-I & $0< \Gamma t \leq \tau_D$ &  ---  & ---   \\
                    & X-Z & ---  & $0 < \Gamma t < 0.25$ & $0.25 \leq \Gamma t \leq \tau_D$ \\
\hline
                   & X-I & ---  & $0.43 < \Gamma t \leq \tau_D$ & $0< \Gamma t \leq 0.43$  \\
$\rho_{W}^{\Phi^{+}}$ & Z-I & --- &  ---  & $0< \Gamma t \leq \tau_D$   \\
                    & X-Z & $0< \Gamma t \leq 0.33$  & --- & $0.33 < \Gamma t \leq \tau_D$ \\
\hline
                   & X-I & $0< \Gamma t \leq 0.33$  & --- & $0.33 < \Gamma t \leq \tau_D$  \\
$\rho_{W}^{\Phi^{-}}$ & Z-I & --- &  ---  & $0< \Gamma t \leq \tau_D$   \\
                    & X-Z & ---  & $0.43 < \Gamma t \leq \tau_D$ & $0< \Gamma t \leq 0.43$ \\
\hline\hline
\end{tabular}
\end{center}
\end{table}
As can be seen from table 5, switching with Z-I gate avoids ESD completely for initial state $\rho_{W}^{\Psi^{+}}$. Figure~\ref{fig:2} shows concurrence evolution of $\rho_{W}^{\Psi^{+}}$ state with and without local switching operations. Concurrence for switching at two different times, namely, $\Gamma t =0.2$ and $\Gamma t =0.28$ is shown in the figure.
ESD can be avoided in the case of initial $\rho_{W}^{\Phi^{+}}$ state when switched with an X-Z gate. Concurrence as a function of time with and without switching for this case is shown in figure~\ref{fig:3}.
It is clear from the table that the switching roles of X-Z and X-I gates on $\rho_{W}^{\Phi^{+}}$ is interchanged in the case of $\rho_{W}^{\Phi^{-}}$. Scheme for avoiding ESD in Werner state can be summarized in the table 6. While there exists a gate operation to avoid ESD in $\rho_{W}^{\Psi^{+}}$ by switching at any time between $0$ and $\tau_D$, there is a small subinterval of $[0,\tau_D]$ for switching with appropriate gate operation to avoid ESD in $\rho_{W}^{\Phi^{\pm}}$.

\begin{table}[h!]
\caption{\label{tabone}Scheme for avoiding ESD in Werner states with $S_{L}(0)=0.7$}
\begin{center}
\begin{tabular}{c c c}
\hline\hline
 Werner state & Time interval & Quantum gate \\
\hline
$\rho_{W}^{\Psi^{+}}$ & (0,$\tau_D$] &  Z-I  \\
$\rho_{W}^{\Phi^{+}}$ & (0,0.33] &  X-Z \\
$\rho_{W}^{\Phi^{-}}$ & (0,0.33] & X-I \\
\hline\hline
\end{tabular}
\end{center}
\end{table}

It is known that state $\rho_{W}^{\Psi^{+}}$ ($\rho_{W}^{\Phi^{\pm}}$) belong to the class of X state with one photon (two photon) coherence. To understand more clearly it is essential to investigate the effect of switching on a class of X states with one photon coherence and two photon coherence.

\subsection{One photon and Two photon coherence states}
In our previous work ~\cite{Namitha2018}, we constructed two classes of X states; one with only one photon coherence and the other with only two photon coherence and investigated the role of initial coherence on entanglement evolution. The class of X states with one photon coherence is given by
\begin{equation}\label{s}
X_1=\left\{\left(
  \begin{array}{cccc}
    \frac{1}{6} & 0 & 0 & 0 \\
    0 & \frac{1}{6} & \frac{x}{6} & 0 \\
    0 & \frac{x}{6} & \frac{1}{2} & 0 \\
   0 & 0 & 0 & \frac{1}{6} \\
  \end{array}
\right);\  0\leq x \leq1.73\right\},
\end{equation}
and the class of X states with two photon coherence is given by
\begin{equation}\label{t}
X_2=\left\{\left(
  \begin{array}{cccc}
    \frac{1}{2} & 0 & 0 & \frac{x}{6} \\
    0 & \frac{1}{6} & 0 & 0 \\
    0 & 0 & \frac{1}{6} & 0 \\
    \frac{x}{6} & 0 & 0 & \frac{1}{6} \\
  \end{array}
\right);\ 0 \leq x \leq1.73\right\}.
\end{equation}
Coherence and concurrence of the states in both classes are same and the analytical expressions obtained are given by $x/3$ and ($\vert x \vert -1)/3$ respectively. We find that the states are entangled for $1 < x \leq 1.73$. We have observed that all the initial mixed entangled states belong to class of states in (\ref{s}) and (\ref{t}) undergoes ESD followed by a revival ~\cite{Namitha2018}. We have also found that the states with high two photon coherence are more entangled and robust towards noise compared to one photon coherence state. We consider the dynamics of these states for a fixed coherence parameter say, $x=1.6$ in (\ref{s},\ref{t}) with $C(0)=0.2$. Before switching, one photon coherence state undergoes ESD at $\tau_D=0.2152$ followed by a revival at $\tau_R=2.9771$ and two photon coherence state undergoes ESD at $\tau_D=0.7467$ with a revival at $\tau_R=1.9854$. Two photon coherence state is shown to have higher ESD time than one photon coherence state and it is discussed in our previous paper ~\cite{Namitha2018}.

\begin{figure}[!h]
\center
\includegraphics[width=15cm, height=10cm]{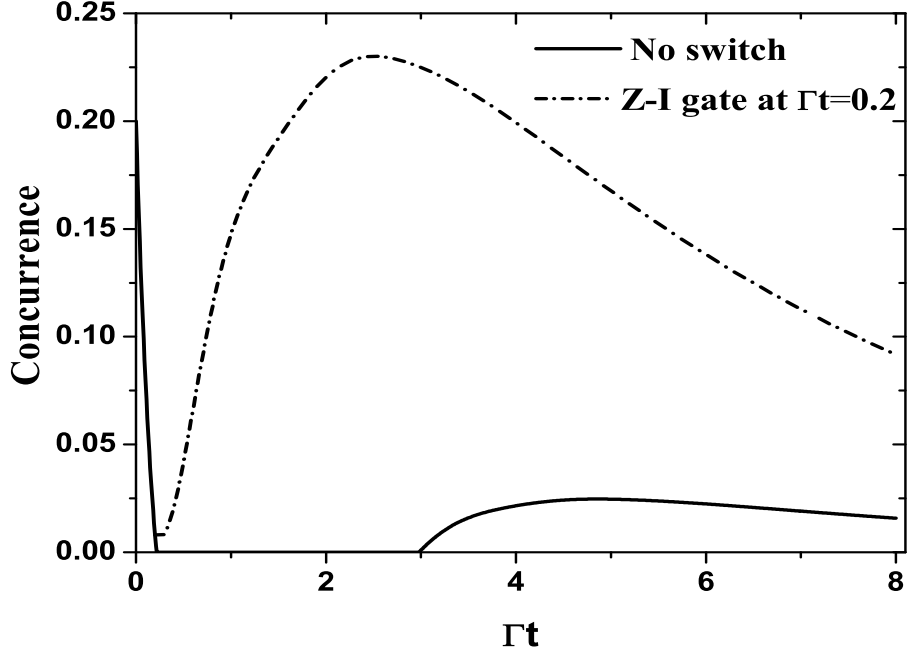}
\caption{Variation of concurrence with time for the evolution of one photon coherence state before (solid black) and after switching with Z-I gate at $\Gamma t =0.2$ (dash dotted red) for $x=1.6$ and $r_{12}=\lambda/6$.}
\label{fig:4}       
\end{figure}

Now, we investigate the effect of switching via different quantum gates on one photon as well as two photon coherence states for a fixed $x=1.6$ with $C(0)=0.2$. First, we investigate the action of switching on one photon coherence via X-I, Z-I and X-Z gates. The effect of switching on ESD time at different switching time is presented in table 7. It is observed that switching with Z-I gate between times $0< \Gamma t \leq \tau_D$, avoids ESD completely and the concurrence in (\ref{C2}) is dominated in the entire the evolution. The concurrence evolution with no switching and with Z-I gate switching at $\Gamma t =0.2$ is shown in figure \ref{fig:4}. It is clearly seen that the switching with Z-I gate avoids ESD completely.
\begin{table}[h!]
\caption{\label{tabone}Switching action in one photon coherence state, $X_{1}$ with $x=1.6$ and $C(0)=0.2$.}
\begin{center}
\begin{tabular}{c c c c c}
\hline\hline
 Initial state & Quantum gates & Avoids ESD & Delays ESD & Hastens ESD \\
\hline
                    & X-I &  ---  & $0< \Gamma t \leq \tau_D$ & --- \\
$X_{1}$ & Z-I & $0< \Gamma t \leq \tau_D$ &  ---  & ---   \\
                    & X-Z & ---  & $0< \Gamma t \leq \tau_D$ & --- \\
\hline\hline
\end{tabular}
\end{center}
\end{table}

\begin{figure}[!h]
\center
\includegraphics[width=15cm, height=10cm]{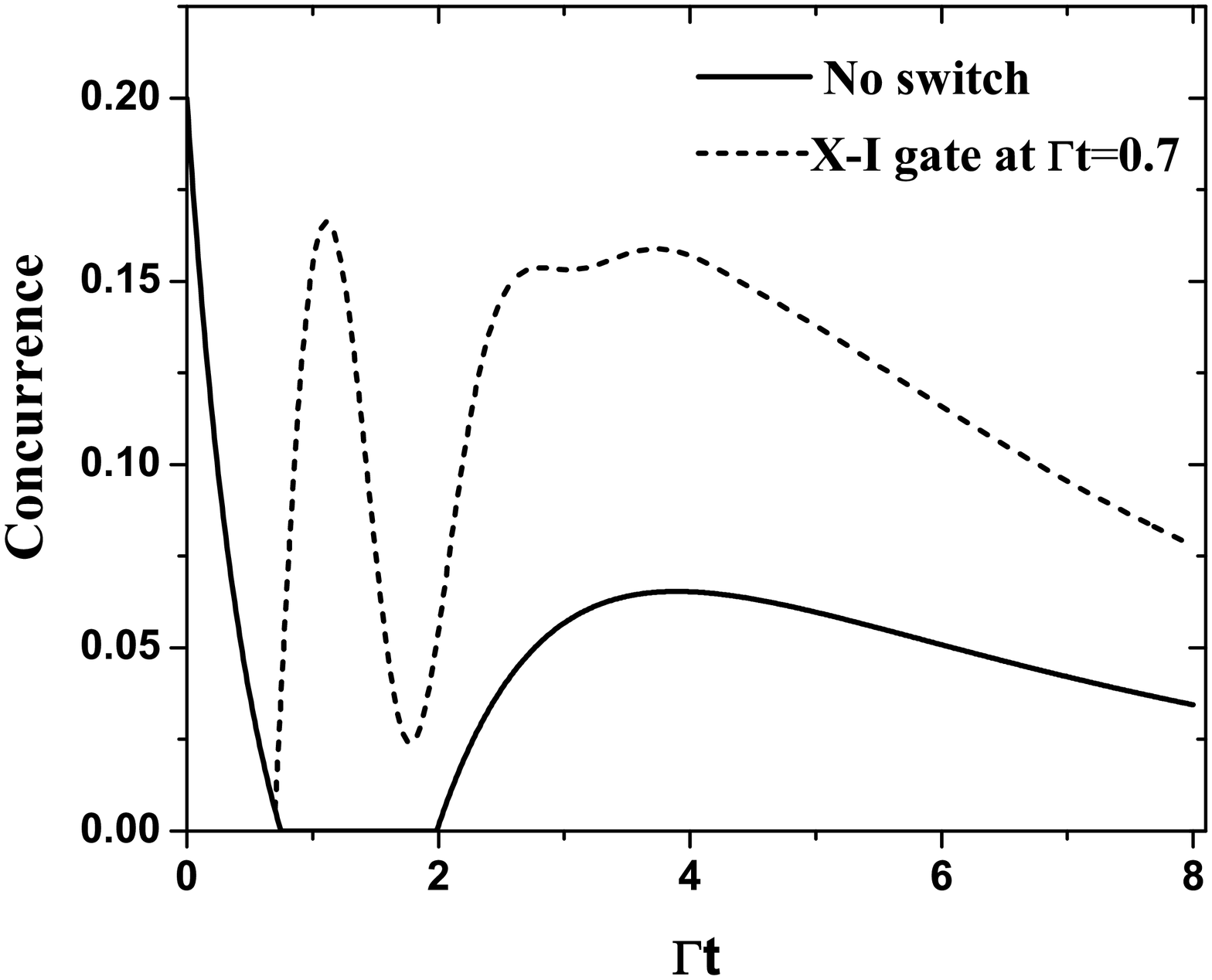}
\caption{Variation of concurrence with time for the evolution of two photon coherence state before (solid black) and after switching with X-I gate $\Gamma t =0.7$ (dashed red) for $x=1.6$ and $r_{12}=\lambda/6$.}
\label{fig:5}       
\end{figure}

\begin{figure}[!h]
\center
\includegraphics[width=15cm, height=10cm]{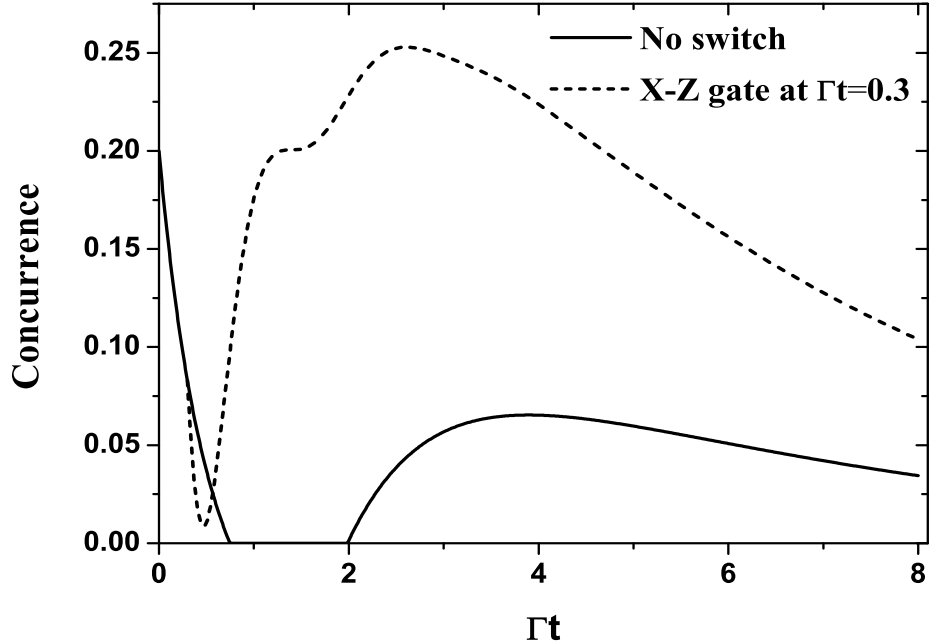}
\caption{Variation of concurrence with time for the evolution of two photon coherence state before (solid black) and after switching with X-Z gate at $\Gamma t =0.3$ (dashed red) for $x=1.6$ and $r_{12}=\lambda/6$.}
\label{fig:6}       
\end{figure}

Now, let us consider the effect of switching on two photon coherence state, $X_{2}$ (\ref{t}) for the same coherence parameter, $x=1.6$ and $C(0)=0.2$. Before switching, the entanglement dynamics till the ESD time is governed by $C_{1}(t)$ and after entanglement revival by $C_{2}(t)$. The entanglement dynamics of two photon coherence state by switching with X-I, Z-I and X-Z gates are systematically investigated and the results are summarized in table 8. The switching of X-I gate causes the state to evolve with the concurrence $C_{2}(t)$ alone and switching between $0.64 < \Gamma t \leq \tau_D$ avoids ESD completely. Figure \ref{fig:5} shows the concurrence evolution of $X_{2}$ state with no switching and with X-I gate switching at $\Gamma t =0.7$. It can be seen from figure that this switching avoids ESD completely. Further, switching with X-Z gate at time between $0< \Gamma t \leq 0.31$ also evolve with concurrence in (\ref{C2}) and found to avoid ESD from the evolution. It is shown in figure \ref{fig:6}. Hence, after switching with X-I and X-Z gate operation $C_{2}(t)$ become dominant term in the calculation of concurrence through out the evolution. Here, $C_{2}(t)$ becomes positive and avoids ESD only if it follows the inequality $\{[\rho_{ss}(t)-\rho_{aa}(t)]^{2}+[2 Im \rho_{sa}(t)]^{2}\}^{1/2} > 2 \sqrt{\rho_{ee}(t)\rho_{gg}(t)}$. We could find out that at what time intervals of switching makes the state to follow this inequality and there by avoids ESD completely and is given in table 8. The switching basically increases the coherence terms $ \rho_{sa}(t)$ which plays a significant role in making $C_{2}(t)$ positive.
\begin{table}[h!]
\caption{\label{tabone}Switching action in two photon coherence state, $X_{2}$ with $x=1.6$ and $C(0)=0.2$.}
\begin{center}
\begin{tabular}{c c c c c}
\hline\hline
 Initial state & Quantum gates & Avoids ESD & Delays ESD & Hastens ESD \\
\hline
                    & X-I & $0.64 < \Gamma t \leq \tau_D$  & $0.28 \leq \Gamma t \leq 0.64 $ & $0< \Gamma t < 0.28$  \\
$X_{2}$ & Z-I & --- &  $0 < \Gamma t \leq \tau_D$  & ---   \\
                    & X-Z & $0< \Gamma t \leq 0.31$  & --- & $0.31 < \Gamma t \leq \tau_D$ \\
\hline\hline
\end{tabular}
\end{center}
\end{table}

In order to avoid ESD for states that belong to the class of one photon coherence states, we have to switch with Z-I gate at any time before ESD time. Other gate operations can delay the ESD and result in faster revival and there by reduces the disentanglement time $\tau_R-\tau_D$. Similarly, we found that the class of states with two photon coherence also can avoid ESD in the process of switching. Switching at early times via X-Z gate and later times via X-I gate can avoid ESD in the two photon coherence states. In the process of switching, we observe that the entanglement produced by one photon coherence in (\ref{C2}) plays a significant role in avoiding the ESD.

\section{Conclusions}
We investigated the effect of local unitary operations on the entanglement dynamics of an interacting two qubit system in contact with a common vacuum bath. Our aim has been to identify schemes involving local unitary switching operations that avoid entanglement sudden death in a select but important class of initial entangled states. We found that for Bell states ($|\Phi^{\pm}\rangle$), ESD can be completely avoided by switching with an appropriate gate among X-I and X-Z gates. The choice of quantum gate used depends on the time of switching. ESD can be avoided by performing switching operation at any time before ESD time (see table 3). We observed that ESD can be completely avoided for specific initial Werner state $\rho_{W}^{\Psi^{+}}$, corresponding to $S_{L}(0)=0.7$ by switching with Z-I gate at any time before ESD time. On the other hand, for $\rho_{W}^{\Phi^{+}}$ ($\rho_{W}^{\Phi^{-}}$) ESD can be avoided if we perform switching operation early on (almost before half of ESD time) using X-Z (X-I) gates (see table 6). Generalization of $\rho_{W}^{\Phi^{+}}$ and $\rho_{W}^{\Phi^{\pm}}$ are a class of one and two photon coherence states. Representative example states of these classes of states show that ESD can be avoided for one photon coherent state by switching with Z-I gate at any time before ESD time, where as for two photon coherent states ESD can be avoided if switching is implemented with X-Z gate early in time and with X-I gate when the time is closer to ESD time. We demonstrated that for important class of initial states, ESD can be completely avoided by choosing appropriate local quantum gate operation and implementing at an appropriate time.

\section*{Acknowledgements}
Namitha C V acknowledge DST for the financial support (No.DST/INSPIRE Fellowship/2013/974) under the scheme of AORC-INSPIRE Fellowship.






\end{document}